\documentclass[fleqn,12pt,twoside]{article}
\usepackage{espcrc1}

\usepackage{graphicx}
\usepackage[figuresright]{rotating}

\newcommand{\AmS}{{\protect\the\textfont2
  A\kern-.1667em\lower.5ex\hbox{M}\kern-.125emS}}

\title{  FEW   BODY   RESERCH - SUMMARY}

\author{I. \v Slaus\address{Rudjer Boskovic Institute, 10001 Zagreb, Croatia, and\\ Triangle Universities Nuclear Laboratory, Durham, N.C. 27708, U.S.A.}
}

\begin{document}

\maketitle

\begin{abstract}
Few-body research history, achievements, current development and challenges are presented.
\end{abstract}

\section{INTRODUCTION}

The first Few Body (FB) conference was held in London in 1959, second in Brela (1967) and---after conferences in Birmingham (1969) and Budapest (1971)---Los Angeles conference in 1972 started a series of conferences held every 2 - 3 years. European FB conferences are held regularly since 1971. In 1975 European FB community was organized, followed by the FB research group of the American Physical Society and recently by Asian-Pacific FB conferences. Few Body Systems - a journal devoted to FB problems - was initiated by W. Plessas in 1986. FB conferences are attended by about 250 - 350 participants from over 40 countries. FB research is more international involving over 20 institutions from 10 countries (and more, e.g. P*ANDA involves 370 physicists from 47 institutions in 16 countries \cite{Bri06}). Through its variety FB field constantly attracts young researchers. While initial conferences were devoted to nuclear and particle physics, today they include particle, nuclear, atomic, molecular, solid state physics, chemistry and astrophysics.

Why did it all start? In 1953 Bethe estimated that more man-hours was devoted to solving nuclear force problem than to any other scientific problem in history \cite{Bet53}. Yet, in 1960 ``scarcely physics owed so little to so many'' \cite{Gol60}. Half a century ago the Los Alamos group \cite{Sea53} and then Zagreb group \cite{Ila61} did many few nucleon studies culminating in the first determination of the neutron-neutron $^1$S$_0$ scattering length, $a_{nn}$, from $^2H$(n,p)nn and $^3H$(n,d)nn yielding $-$21 $\pm$ 1 fm and $-$18 $\pm$ 3 fm \cite{Ajd65}, respectively, with no theoretical uncertainties included. FB studies yielded nuclear data necessary for nuclear energy, e.g. fusion \cite{Lle05}. Kinematically complete (exclusive) measurements by the Rice \cite{Phi60} and BNL \cite{Zup65} groups demonstrated quasifree scattering (QFS) and sequential decay - final state interactions (FSI). UCLA, Caltech, Berkeley and other groups studied $^2H$(p,pp)n, nucleon-nucleon (NN) Bremsstrahlung (NN$\gamma$) and excited states of $^4$He. 4$\pi$ measurements were initiated by the IKO group developing BOL detector system \cite{Wie71} and by the Zagreb group using nuclear emulsions \cite{Ant72}.

Discoveries of heavier mesons in the early 60-ies gave birth to one-boson-exchange (OBE) potentials \cite{Mac89}.The founding fathers of nuclear physics were aware of the importance of three nucleon forces (3NF): ``the description of nuclei in terms of NN force except for the deuteron cannot be considered satisfactory''. The number of m-body interactions in an n-body system is n!/m!(n-m)! and that is already for $^6$Li larger than n!/2!(n-2)! and it grows with A \cite{Pri39}. In 1957 the first 3NF was developed \cite{Fuj57}.

Quantum three body problem was solved exactly by L. Faddeev in 1960 \cite{Fad60} followed by numerical results by Mitra \cite{Mit62} and Amado \cite{Ama63}, albeit using S-wave separable potentials. Based on Faddeev theory Sandhas et al \cite{Alt67} in 1967 developed the AGS equation - our basis ever since.

During this half a century a remarkable progress was achieved in nuclear forces, in three body calculations and in experiments. New ideas, new approaches and surprises marked this period extending the breadth of the FB field. I will attempt to give glimpses of this remarkable story which is being told so eloquently by each of you by your own contributions to this successful endeavor.

\section{PARADIGM CHANGE IN THE NUCLEON-NUCLEON STUDIES}

Current NN potentials form two groups, see Table 1.  
1) High precision potentials \cite{Shi06,Dol05,Kuk02,Mac01,Wir95} fit about 6000 NN data up to 350 MeV with a $\chi^2$/datum about 1.0, comparable to the phase shift analysis (PSA) \cite{Sto93}. These potentials need 30-40 parameters. The CD-Bonn potential, based on OBE model, is charge dependent (CD) and it reproduces pp data below 350 MeV with $\chi^2$/datum = 1.01. AV18 includes electromagnetic (EM) forces, one-pion-exchange (OPE) and short term phenomenology. EM includes one gamma and two gamma exchanges, vacuum polarization, Darwin-Foldy and magnetic interaction (MI). All can be written as a sum of 18 operators  (thus AV18). Potentials from inverse scattering (JISP) \cite{Shi06} adjust their off-shell behavior to mimic 3NF. JISP6 and JISP16, where numbers 6 and 16 indicate whether off shell has been fixed by fitting binding energies (BE) of A=6 or A=16 nuclei, lead to rapid convergence in the No-core Shell Model (NCSM) \cite{Nav03}. The composite structure of hadrons and relativistic dynamics lead to nonlocalities at short range and Doleschall's potential is non-local at r$\leq$3 fm with OPE tail \cite{Dol05}. Kukulin's potential \cite{Kuk02} is a realization of a hybrid quark-meson model. Several 3NF are used: Tucson-Melbourne (TM) \cite{Coo79}, TM99' \cite{Coo01} improved to be consistent with chiral symmetry, Urbana/Illinois: UIX \cite{Pud95} and IL1-5 \cite{Pie01}. An alternative mechanism of generating 3NF uses explicitly $\Delta$-isobar in a coupled-channel approach\cite{Haj83}. The force employed in the isobar approach, referred as CD-Bonn + $\Delta$, is purely nucleonic in T=0 states and the coupled-channel two-baryon acts in T=1 states where few constants in CD Bonn are retuned. It reproduces NN data with $\chi^2$/datum of 1.02.
2) NN, 3N, 4N.. potentials based on chiral perturbation theory ($\chi$PT). The QCD Langrangian L for massless u and d is chirally symmetric. The effective L starts with the most general L consistent with all symmetries particularly broken chiral symmetry of QCD. At low energies effective degrees of freedom are nucleons and pions, while heavy mesons and deltas are “integrated out”.  In the chiral effective field theory (EFT) the contributions are made up by a series in (Q/$ \Lambda$)$^n$, Q= momentum or pion mass,  $\Lambda$ = 1 GeV, n$\geq$0. For a given n the number of contributing NN, 3N, 4NF diagrams is finite and calculable. Uncertainty at order n can be estimated by calculating the next order. Parameters in $\chi$PT are constrained by $\pi$N and $\pi\pi$ scatterings and consistent among NN, 3N, 4N... Table 1 shows that N3LO is needed to get a good $\chi^2$/datum. $\chi$PT generates NN, 3N, 4N etc from the same effective L. The 3NF appears for the first time in N2LO explaining why 3NF is weaker than the NN force. Due to partial cancellations 3NF becomes noticeable, e.g. in BE of $^3$H 0.5-1.0 MeV out of 8.5 MeV, of $^6$Li 6 out of 32 MeV and in scattering processes, see V.2 and VII. The 3NF that appears in N2LO differs from UIX and IL 3NF, which consists of chiral terms of higher order. $\chi$PT is fundamental and quantitative and its appropriate use is equivalent to using QCD. Though $\chi$PT can be carried to an arbitrarily high order, predictive power is limited due to increase in the number of parameters \cite{Mei05,Epe05,Ent03}.
The excellent description of NN data and the promising procedure ($\chi$PT EFT) are the work of few groups led by J.J. de Swart, R. Machleidt, V. Pandharipande, P.U. Sauer and U-G. Meissner.

\section{PARADIGM CHANGE IN FEW BODY THEORIES}

Only approximate calculations of the three body equation existed until Glöckle and collaborators achieved a breakthrough - rigorous three body calculation (r3Bc)\cite{Glo96} employing high precision NN forces, including TM, TM99', UIX and ILi, and now the Coulomb force \cite{Alt78,Del05,Kie04,Viv06} and NnLO forces. Benchmark studies proved that various r3Bc are consistent within 1\%. The r3Bc in most cases predicts the data very well. Of course, EM forces have to be included also at high energies (130 MeV) when breakup configurations include small angles (15$^\circ$) \cite{Kys06}. We can use r3Bc in planning experiments. Extensive studies at 13, 65, 135 and 200 MeV on sensitivity to NN and 3NF were done \cite{Kur02}. Sensitivity increases with incident energy. Some observables are very sensitive, e.g. at 200 MeV tensor analyzing powers up to 400\%. These areas of high sensitivity are concentrated in specific regions and it is a credit to early researchers who intuitively or experimentally established them, e.g. symmetric constant relative energies (SCRE) (among them space star), collinearity \cite{Lam76}, FSI and QFS (in early 70-ies Tjon \cite{Wie76} studied sensitivities with S-wave potential). Several different groups are doing r3Bc \cite{Sus04,Ish03,Dol05,Bra06}.

FB studies include also systems with four and more bodies. Faddeev theory was extended by Yakubovsky \cite{Yak67}, Sandhas' group \cite{Gra67}, Fonseca \cite{Fon99}, Pisa group \cite{Viv01} and by developing NCSM and Green Function Monte Carlo (GFMC) \cite{Pie05}. Richness and complexity increases as A increases. A=4 exhibit BE saturation, excited states and rearrangement processes. Hep process is important for solar models and Big Bang Nucleosynthesis (BBN)\cite{Laz05}. Results of several calculations based on Faddeev theory, GFMC and NCSM are listed in Table 2 and a remarkable agreement is obvious. AV18+UIX still underbinds nuclei starting with $^6$He and the discrepancy increases with A. Fitting energies of 17 states in A$\leq$8 new 3NF were constructed: IL 1-5 which contains besides 2PE also 3PE and $\Delta$s. 3NF potentials IL1-5 are CD. Comparison with 17 states and AV18 gives average deviation of 7.32 MeV. When the 3NF is included, the average deviation decreases to 2.02 MeV for AV18+UIX and to 0.04 MeV for AV18+IL3. Most significant is $^{10}$B, where AV18 predicted two 1$^+$ states below the experimental 3$^+$ state, AV18+IL2 predicts correct ordering and correct energies. Similarly, for $^9$Be IL2 is causing an inversion of levels bringing them into agreement with data. This is due to an increased spin-orbit in IL2. The NCSM also gives correct level ordering for $^{10,11,12}$B and $^{12}$N when the 3NF is included \cite{Cou06}.
3B theory was applied to complex nuclear processes \cite{Haf77} and to hypernuclei e.g. to low lying levels of $^9_\Lambda$Be treating it as the 3B system: $ \Lambda\alpha\alpha$ \cite{Fil06}.

\section{TECHNOLOGY WE DEVELOP CHANGES OUR
RESEARCH AND US}

Developing and building instruments played an important role throughout the history of science. During this half a century major developments occurred in: accelerators, detectors, electronics, computers and data analysis. Few highlights follow. Review of facilities in Europe, the USA, Russia and Japan is given in ref\cite{Few04}.
There are now 17,500 accelerators: 120 producing beams of energies larger than 1 GeV, about 1,000 low energy accelerators for research, 100 synchrotron radiation facilities, more than 7,500 radiotherapy accelerators (mainly linacs) and more than 7,000 for ion implantation and condensed matter work \cite{Ama05}. Tandems played important role in our research in 1960-90-ies and few, notably TUNL and Cologne, are still leading centers. Early SF cyclotrons, then LAMPF, TRIUMF, PSI, IUCF, RCNP, MIT/BATES and now JLab (to be upgraded to 12 GeV), KVI, HI$\gamma$S (High Intensity Gamma Ray Source at Duke), MAMI -Mainz Microtron to be upgraded to 1.5 GeV, CELSIUS - accelerator and storage ring at Uppsala, LEGS - Laser electron gamma source of highly polarized monochromatic gamma 150 - 470 MeV by Compton backscattering from 2.8 GeV electron beam at National Synchrotron Light Source at BNL, ESRF - European Synchrotron Radiation Facility, storage ring at Uppsala and electron storage ring TERAS at AIST, Tokyo are playing a crucial role in our research. Radioactive ions beams (RIB) opened new insights. For the first time it is possible to use beams of N/Z = 3, e.g. $^8$He facilitating the transfer of 4n. RIB are produced by: i) isotope separation on line using two independent accelerators: accelerator for production of short lived radio-nuclei in a thick target connected to an ion source and a post accelerator, ii) in flight which relies on very energetic beams of heavy ions impinging on a thin target and resulting in fission or fragmentation with velocities of outgoing particles comparable to velocity of the incident projectile. Cocktail of different particles identified by mass, charge and momentum are produced and do not need further acceleration. And iii) a combination of i) and ii). A variety of RIB are available with very short lifetimes, e.g. $^{14}$Be (T= 4.2ms), $^8$He (T= 119ms), $^6$He (T= 0.807s), $^8$Li (T=0.838s) and “long-lived” $^{10}$C (T=19.3s) and $^{18}$Ne (T=17s). High beam intensities are achieved: up to billion particles per second. The number of RIB facilities is increasing. Not a complete list includes ISOLDE at CERN, SPIRAL at Ganil, FAIR (construction begins in 2008) at GSI, ISAC at TRIUMF, RIBF at RIKEN (under construction) and SBL (both in Japan), NSCL at MSU, Louvain-la-neuve, Orsay, ORNL, Berkeley, FSU, RIBRAS (Sao Paolo), RIBLL (Lanzhou, China), Argonne and Catania \cite{She05}. DOE is considering preliminary engineering design of advanced exotic beam facility in 2011.
Several kaon beams were developed strengthening hypernuclei research: e.g. COSY, JLab, Nuclotron at Dubna and DA$\Phi$NE (Double Annual ring For Nice Experiments), Frascati, colliding e$^-$  and e$^+$ beams to produce $\Phi$(1020) decaying into neutral and charged kaons. DA$\Phi$NE is delivering about 150 $\Phi$/s. Beams of antiparticles, e.g. antiprotons and antideuterons are developed.

Detectors range from single to complex structures matching in R\&D and cost those of accelerators. Detectors are used in stand-alone mode as in underground neutrino experiments or with accelerators. A stand-alone detector is the proposed Majorana $^{76}$Ge double-beta decay detector: a 500 kg of $^{76}$Ge isotopically enriched to 86\% allowing a sensitivity of half-life measurement of $^{76}$Ge of 4$\times$10$^{27}$ y over a data acquisition period of 5000 kg years \cite{Aal05}. Some examples of detectors used with accelerators are: JLab CLAS, SALAD-KVI (small angle large acceptance detector) (AGOR facility), now BINA, RIKEN spectrograph SMART; Gammasphere array of 100 large Compton-suppressed Ge crystals, GRETA and GRETINA (gamma ray energy tracking in-beam nuclear array), Crystal Ball (at AGS and now at MAMI), Crystal Barrel and TAPS at ELSA, WASA (wide angle shower apparatus) at CELSIUS and multidetector systems developed at Erlangen, TUNL and Cologne, superconducting magnetic spectrograph S800 at MSU, Blowfish detector array consisting of 88 neutron detectors, ANKE detection system consisting of range telescopes, scintillation counters and MWPC at COSY and detection systems developed at JLab. Detection systems for hypernuclear research are hyperball, FINUDA and P*ANDA.

Astronomical observations require measurements of high energy gammas up to 10$^{20}$ eV, rapidly varying gammas in TeV range, and neutrinos. Underground neutrino detection system include Kamiokande, SuperKamiokande, BAKSAN, SNO, KamLAND and a next generation of solar neutrino detectors BOREXINO in Gran Sasso, detectors under the South Pole (Amanda and Ice Cube), underwater detector projects in the Mediterranean (off the coasts of France and Greece) and in Russia.

Computers - for high speed complex computing, for automated measurement and control and for storage, retrieval and data analyses - coupled with networking, internet, www and fusing computers and telecommunications - revolutionized research. Higher energies and higher intensities of accelerators open the possibility for processes generating vast amount of information that has to be stored and analyzed and total accumulated data in a typical experiment are now orders of magnitude larger than few decades ago. Software cuts on kinematical conditions and particles ID reduce the data stored on tape by several orders of magnitude. Many studies would be impossible without Monte Carlo simulations. The progress from tube computers that were used in the 50-ies with punched cards to present supercomputer, DNA computer constructed in 2004 by E. Shapiro at Weizmann Institute and a prospect for quantum computing made a paradigmatic change.

\section{FEW BODY RESEARCH - RESULTS AND PROBLEMS}

There is an impressive amount of FB data. NN and nucleon-deuteron (Nd): proton-deuteron (pd) and neutron-deuteron (nd) elastic and breakup data: total, partial and differential cross sections, spin observables measurements involving N and d. The longstanding issue of $a_{nn}$ will be addressed by doing nn scattering experiment (first tried with underground nuclear explosion) using the pulsed aperiodic Yaguar reactor at Snezhinsk giving neutron flux of density of over 10$^{18}$ cm$^{-2}$ s$^{-1}$ \cite{Fur02}.
Some results and problems:

V.1. Symmetries:

V.1.1. Unitarity of the Cabibbo - Kobayashi - Maskawa matrix \cite{Yao06}. The Particle Data Group reported in 2002 $V_{ud} = 0.9734\pm 0.0008$ (from beta and neutron decays),
V$_{us}= 0.2196\pm0.0023$ (from K$^+$ and K$^0$ decay) and $V_{ub}= 0.0036\pm0.0007$ (from semileptonic B decay), wherefrom breaking of unitarity occurred:
$Delta = 1 - (|V_{ud}|^2 + |V_{us}|^2 + |V_{ub}|^2) = 0.0043\pm 0.0019$, i.e. 2.2 $\sigma$ \cite{Har05}.
	Several measurements done after 2002: BNL E865, NA 48, KLOE, KTev and ISTRA+ including K and hyperon decays gave $V_{us} = 0.2261\pm 0.0021$ leading to $Delta = 0.0004\pm 0.0011$ compatible with unitarity \cite{Buc06}. Studies of pion decay and of 12 superallowed transitions $0^+ \rightarrow 0^+$ from $^{10}$C to $^{87}$Rb are currently performed \cite{Har04}.

V.1.2. CPT invariance \cite{Yao06} requires the equality of masses and lifetimes of particles and antiparticles. The upper limit of 10$^{-18}$ is established based on the equality of masses of K$^\circ$ and its antiparticle.

V.1.3. CP violation \cite{Yao06} in the quark sector is known since 1964 and research goes on  (evidence in $B^\pm \rightarrow K\pi\pi$; permanent electric dipole moment (EDM) of any fundamental particle violates P and T, limits on EDM for electron is 1.6 $\times$ 10$^{-27}$ ecm, for neutron 6.3 $\times$ 10$^{-26}$ ecm and for proton $-$3.7$\pm$6.3 $\times$ 10$^{-23}$). However, nothing is known about CP violation in the lepton sector. Based on solar and atmospheric neutrino and reactor antineutrino measurements neutrinos do have mass (otherwise there would be no oscillation, albeit according to WMAP observations the sum of masses of three neutrino flavors is less than 1 eV \cite{Ben03}). Comparison of behavior of neutrinos and antineutrino provides information on CP violation in the lepton sector.

V.1.4. Conservation of lepton number: neutrinoless double beta decay $(Z,A) \rightarrow (Z+2,A) +e^- +e^-$
studies yield for the upper value of the $^{76}$Ge lifetime 1.9$\times$10$^{25}$ y with a confidence limit of 90\% \cite{Yao06}.

V.1.5. Time variation of the fine structure constant $\alpha_{QED}$: observation of the variability of basic constants could give information on the existence of extra dimensions \cite{Uza03}. BBN offers opportunity to test time variation of fundamental constants since reaction rates depend on fundamental constants \cite{BLi06}. Crucial rate is that for $n p \rightarrow d \gamma$ at 30 - 130 keV. The 5\% uncertainty associated with this process contributes a significant uncertainties in the abundances of light elements produced in the BBN. This process was measured at HI$\gamma$S and the data \cite{Tor03} are in agreement with the EFT calculation  \cite{Che99} and with the NN potential model of ref \cite{Are91} thereby reducing the uncertainty of 5\%. Discovery that the quasar absorption lines in interstellar media suggest a small evolution of the $\alpha_{QED}$: $\dot{\alpha}/\alpha$ = (6.40$\pm$1.35)$\times$10$^{-16}$/year \cite{Web99} is related to the predictions of string theory and Kaluza-Klein theory \cite{Pol98}. This observation is not confirmed. Laboratory experiments using narrow atomic hydrogen 1s $\rightarrow$ 2s transition compared to atomic clock did not see any change \cite{Har04}. Primordial helium abundance depending on 6 fundamental constants and WMAP observations put a bound on $\Delta\dot{\alpha}/\alpha$ of few times 10$^{-3}$. Stringent bounds can be obtained from deuterium and lithium primordial abundances \cite{Mue04}.

V.1.6. Supersymmetric particles search \cite{Yao06} resulted in lower limits for their masses, e.g. for neutralino 46 GeV, for chargino 94 GeV, for selectron 73 GeV, for squark 259 GeV and for gluino 195 GeV with CL of 95\%.
and

V.1.7. Charge dependence and charge symmetry breaking (CSB)
	These were the first broken symmetries studied resulting in SU(2) and SU(3) and non-Abelian gauge theories. On the fundamental level CD and CSB are due to differences between masses and charges of u and d quarks \cite{Mil90}. The consequences on the hadronic level are: i) the difference among hadron masses: m$_d$ - m$_u$ manifests itself in $\pi^\pm$ - $\pi^0$, and n - p mass differences (m$_n$  $>$ m$_p$ has significant consequences for the structure of the Universe), ii) meson mixing, $\rho$-$\omega$ is important, and iii) irreducible meson-photon exchanges. Pion mass difference accounts for 60\% of CD in $^1$S$_0$ \cite{Mil90}. Nucleon mass splitting has several consequences. The most trivial is the change in the kinetic energy whereby the $a_{nn}$ becomes more negative by 0.25 fm. OBE diagrams are negligibly affected, but TPE are. The $\rho$-$\omega$ mixing effect was subject to criticism and counterarguments and this issue is still open \cite{Pie93}.
CSB is observed in: i) $\delta a_{exp} = a_{pp} - a_{nn} = (- 17.3 \pm 0.3) - (- 18.9 \pm 0.4) = 1.6 \pm 0.5$ fm  (Only the $a_{nn}$ value from the $\pi^-$d measurement is taken -18.93$\pm$ 0.27$_{ex}$ $\pm$0.3$_{th}$ \cite{Che06}. Reassessing theoretical uncertainties \cite{Gar06} leads to even smaller value of $\pm$0.2$_{th}$. Values from nd breakup are discussed in V.3). Various models predict for $a_{pp} - a_{nn}$ values in agreement with this experimental result: chiral quark model \cite{Ent99} predicts 1.37 - 1.67 fm and it correctly predicts $a_{np}$ = $-$23.749 fm. CSB due to $\rho$-$\omega$ mixing gives 1.508 fm, the same result as the TPE taking into account n-p mass difference. Therefore, $a_{pp}-a_{nn}$ cannot distinguish between two hadronic mechanisms generating CSB \cite{MMu01}. ii) $^3$H and $^3$He BE difference is 764 keV,  mainly due to EM effects and after all corrections the 65$\pm$10 keV is attributed to CSB \cite{Nog03}. Isospin violating 3NF within the EFT contributes 5 keV to the CSB in the $^3$H-$^3$He \cite{Fri05}. CSB in $^3$H-$^3$He is due mainly to S-wave forces: CSB potential derived from AV18 fixed by the best fit to pp data, and then adjusting its central force in the S=0, T+1 state to fit $a_{nn}$ gives for  $^3$H-$^3$He 62.1 keV for $^1$S$_0$ and 65.1 keV when all T = 1 states are included\cite{MMu01}. Results for CSB due to $\rho$-$\omega$ mixing and due to n-p mass difference are 60.9 and 57.6, respectively for $^1$S$_0$ and 65.7 and 60 keV when all T = 1 states are included. For the S=1 and T=1 states ($^3$P$_J$) it is assumed that the CSB splitting is the same as in S=0, T=1. iii) Nolen - Schiffer anomaly - energy difference between mirror nuclei that cannot be accounted for by EM and nuclear models - is of the order of 200 - 400 keV and 50\% comes from L$\geq$0 NN forces \cite{MMu01}. Nuclear matter is sensitive to L$\geq$0 forces. CSB AV18 and CSB potentials due to n-p mass difference and $\rho$-$\omega$ mixing predict different $^3$P$_J$ phases: nn and pp $^3$P$_0$ and $^3$P$_1$ differ by 0.5$^\circ$ to 1.0$^\circ$ and $^3$P$_2$ by less than 0.2$^\circ$ \cite{MMu01,Lan82}. CD in $^3$P$_J$ waves was studied through PSA \cite{Sto88} and found that np and pp $^3$P$_0$ at 25 Mev differ by 2.2$^\circ$. iv) difference between n and p analyzing powers studied at 183 MeV \cite{Knu91}, 347 MeV \cite{Zha98} and 477 MeV \cite{Abe86}. Theory predicts data very well \cite{Hol87} and $\rho$-$\omega$ mixing gives different contributions at different energies. v) superratio of $\pi^+\pi^-$ on $^3$H and $^3$He. vi) total cross section for D(d, $\alpha$)$\pi^0$ measured \cite{Ste03} at 228.5 and 231.8 MeV at IUCF and found to be 12.7$\pm$2.2 pb and 15.1$\pm$3.1 pb smaller than simplified model \cite{Gar04} based on the EFT predictions of 23.0 pb and 30.8 pb. vii) asymmetry in $n+ p \rightarrow d + \pi^0$ $\sigma(\theta)$ measured at TRIUMF: (17.2$\pm$8st$\pm$5.5sy)10$^{-4}$ \cite{Opp03}. In the framework of the EFT main contributions are mixing and rescattering which through partial cancellation produce an upper value of 69$\times$10$^-4$ \cite{Kol00}. viii) recent theoretical analysis of the $\pi N$ data found a small CSB and ix) $ \Lambda$ separation energies in $^4_\Lambda$H and $^4_\Lambda$He, which after removing Coulomb effects amounts to 390 keV, i.e. about five times larger than the CSB in A=3.
	Extraction of strangeness form factors, the NuTeV anomaly and hadronic corrections to g-2 are related to CSB \cite{Mil06}. Parity-violating L-R asymmetry in e-p scattering is sensitive to s quark contributions and measurement at JLab\cite{Arm05} has to be corrected by CSB. Ratio of charged-current and neutral-current cross sections for neutrinos and antineutrinos on isoscalar target is equal to $1/2 - \sin^2 \theta_W$. Measurements \cite{Zel02} give sin$^2 \theta_W$ three standard deviations larger that Standard Model (SM) prediction. CSB corrections account for about a half of the discrepancy \cite{Lon03}. $g-2$ is known to 1 in 10$^6$, but it is 1 - 2.4 standard deviations off SM predictions and accurate CSB effects of $\rho$-$\omega$ mixing account for 20\% \cite{Mal05}.

V.2. Evidence for the 3NF:
Composite structure of nucleons and $\chi$PT EFT lead to 3NF. We summarize experimental evidence for 3NF:

1) High precision NN forces underbind $^3$H BE and the disagreement gets worse as A increases. There is a correlation among 3N and 4N BE \cite{Tjo75} recently extended to light nuclei and nuclear matter \cite{Del06a}. Phenomenological 3NF with parameters adjusted to predict 3H BE can account for BE and energy levels up to A = 16 \cite{Coo01,Pie01,Haj83}. Using the same NN and 3NF a fit to $^4$He is obtained \cite{Nog02} indicating that 4NF gives a small contribution. Ordering of energy levels of light nuclei, particularly of $^{10}$B, requires 3NF and IL2 with increased LS gives the best description.

2) PSA of pd elastic scattering E$_p$ = 4 - 22.70 MeV data requires 3NF to achieve compatibility between 3N and NN phase shift \cite{Che04}. Just as AV18 cannot predict Et, it fails with $^2$S$_{1/2}$, and agreement is achieved if UIX is added. Ay and iT$_{11}$ depends on $^4$P$_J$ and these are related to the NN $^3$P$_J$. The correct description of Ay depends on $^2P_{3/2} - ^4P_{3/2}$ mixing $\epsilon_{3/2-}$. Small changes in $\epsilon_{1/2-}$ are required to describe iT$_{11}$ and Ay. Essential problem is that $^4$P$_J$ required by 3N PSA are incompatible with $^3$P$_J$ demanded by the NN PSA. 3NF affecting  $^4$P$_{1/2}$ and $\epsilon_{3/2-}$ are required and it seems that the LS type 3NF from refs \cite{Kie99,Can01} fulfills this role.

3) R3Bc with the CD Bonn \cite{Wit99} fit total nd cross section data \cite{Abf99} up to 150 MeV. Above this energy and up to 200 MeV inclusion of the 3NF fits the data. Relativistic kinematics increases the total cross section, but boost effect outweighs it \cite{Wit05}. Including 3NF and relativity still does not account for the nd total cross section at high energies. Data up to 200 MeV provide evidence for the 3NF.

4) Disagreement between the data at the minimum in the elastic Nd scattering differential cross section and the r3Bc using NN forces is decreased when the 3NF are included \cite{Wit98,Sek04}. Measurement of the nd $\sigma(\theta)$ at 95 MeV \cite{Mer04}covers the full angular distribution (20$^\circ$-160$^\circ$) measuring neutrons and deuterons at forward angles using detector SCANDAL at Uppsala University. Data from 100$^\circ$-160$^\circ$ show the need for the 3NF. Relativistic effects modify $\sigma(\theta)$ in the domain around the minimum and in general increase the cross section therefore having a similar effect as the 3NF.

5) Extensive study of H(d,pp)n at 130 MeV performed at KVI with SALAD system \cite{Kys05} in 72 kinematic configurations comparing the cross section data with NN, NN + 3N, isobar and N2LO (with NN and 3N) predictions, ignoring EM interactions \cite{Wit06}, shows agreement between the theory and of the data, but also some disagreements. Fits to the data allowing for 10\% uncertainty in the overall normalization is improved if 3NF are added. Some discrepancies still remain.

       6) High precision NN potentials could not explain polarization transfer coefficients measured at 22.7 MeV \cite{Wit06}, but a good agreement is obtained when 3NF TM' and UIX are included. NLO does not fit the data, but N2LO which includes NN and 3NF does.

       7) Proton-deuteron elastic scattering observables Ay, iT$_{11}$ and Cyy at 197 MeV incident proton energy cannot be explained by CD Bonn or AV18 NN potentials but fit is obtained adding TM 3NF \cite{Cad01}.

V.3. Experimental evidence not compatible with current theories

1) There are discrepancies between $\pi$NN coupling constants extracted from different observables \cite{Mac00}.

2) Cross sections $^3$He($\gamma$,p)d and $^3$He($\gamma$,pp)n at 10.2 and 16 MeV \cite{Nai06} are compared with AV18 and AV18+UIX. Theory predicts 10-14\% larger cross section at 16 MeV and at 10.2 MeV predicts  $^3$He($\gamma$,pp)n a factor of 3 too high. Neutron energy spectra measured at HI$\gamma$S using 12.8 MeV linearly polarized gammas peak at forward angles at lower energies than r3Bc with CD-Bonn and Coulomb \cite{Wel06}.

3) nd and pd Ay data revealed 25 years ago difference between them and the 3B calculation \cite{Tor82,Gru83}. As the accuracy of the data improved and additional data accumulated and as r3Bc now including EM are made the disagreement became more convincing. Inclusion of MI raises the calculated Ay for pd by 4\% bringing it closer to the data, but decreasing by 3\% for nd increasing the disagreement. MI is appreciable at energies below 5 MeV. EM plays a significant role \cite{Del05,Kie04}, but it does not solve the Ay puzzle at low energy: difference between the nd and pd Ay peak values and r3Bc from very low energies to about 20 MeV \cite{Nei03}. Explaining these disagreements would require a long range 3NF. Success of the LS 3NF \cite{Kie99} and Kukulin's potential \cite{Kuk06} require further studies. A remedy would be CD and CSB in the $^3$P$_J$ waves, since Ay and iT$_{11}$ are a magnifying glass for the $^3$P$_J$ force. Such attempts were made unsuccessfully \cite{Wit91,Sol96}.

4) Though Ay data agree with r3Bc at incident nucleon energies of 30 MeV, the Ay  puzzle reappears at higher energies, i.e. at 150 and 190 MeV, where data are not reproduced by high precision NN + TM \cite{Bie00}. Disagreement is also found in deuteron tensor analyzing powers \cite{Kur02}.

5) SCRE, particularly space stars investigated at incident proton energy of 10.5 - 130 MeV and for 10.3 - 25  MeV neutron energy by several different groups. \cite{Str89,Rau91,Zho01}. Data are consistent: at 10.3 and 13 MeV nd has 50\% larger cross section than pd. At 10.5 and at 13 MeV  r3Bc even including the isobar and EM do not explain either pd or nd data. 3NF and MI have negligible ($\leq$ 2\%) effects and the cross section is dominated by well known S-wave. R3Bc fit well at higher energies.

6) While r3Bc fits well np QFS in a wide range of incident energies, it does not explain nn and highly accurate pp QFS data. Including the Coulomb force brings r3Bc in better agreement with the pp QFS data, MI effects are less than $\leq$ 2\% and effects of the $\Delta$ or other 3NF are small.

7) Four-body calculations with high precision NN and UIX 3NF do not fit the n - $^3$H total cross section data \cite{Laz05}.

8) Studies of nn FSI in nd breakup using r3Bc were done at incident energies 13 - 25 MeV. The $a_{nn}$ influences the absolute cross section and the shape of the FSI enhancement and $a_{nn}$ was extracted by both approaches. Kinematically incomplete study at 17.4 MeV \cite{vWi06} gave $a_{nn}$ = ($-$16.5$\pm$0.69st$\pm$ 052sy) fm. The study at 13 MeV measuring neutrons' time-of-flight at $\theta_{n1}$=$\theta_{n2}$= 20.5$^\circ$, 28$^\circ$, 35$^\circ$ and 43$^\circ$, and the deposited energy of the charged particle gave $a_{nn}$ = ($-$18.72$\pm$0.13stat$\pm$0.65sys) fm \cite{Gon06} using absolute cross section, and ($-$18.84$\pm$0.47) fm using the shape. Earlier studies at 13 MeV \cite{Geb93} did not use the r3Bc, but if its shape is reanalyzed it yields $-$17.9$\pm$0.5 fm. Studies at 16.6 MeV and 25.3 MeV measuring the momenta of the outgoing p at 41.2$^\circ$ and n at 55.5$^\circ$ at opposite sides of the beam and consequently using a thin deuterium target gave \cite{Huh00} $a_{nn}$ = ($-$16.2$\pm$0.3) fm and ($-$16.3$\pm$04) fm, respectively. This disagreement between $a_{nn}$ extracted at 13, 16.6, 17.4 and 25.3 MeV is not understood. With additional neutron detectors at 55.5$^\circ$, 69$^\circ$ and 83.5$^\circ$ at the opposite side of the beam np FSI was studied giving $a_{np}$ = $-$23.5$\pm$0.8 fm \cite{Gon99}. In a follow-up study at 25.2 MeV measuring neutron and proton at the same angle of 32o gave $a_{np}$ = -24.3$\pm$1.1 fm \cite{Huh00}. These comparisons validate both studies to an extent that r3Bc explains np FSI. However, it is known that some np FSI configurations are not well described by the r3Bc. Joint effort by both groups is in progress at 19 MeV using simultaneously both geometries: 1) n-n coincidences measured at $\theta_{n1}$=$\theta_{n2}$= 35$^\circ$ and 51.7$^\circ$ and 2) n - p coincidences at $\theta_p$ = 45$^\circ$ and $\theta_n$ = 51.7$^\circ$. Preliminary result for the first geometry is (-17.6$\pm$0.2st$\pm$0.9syst) fm \cite{Cro06}. Discrepancy between results for $a_{nn}$ could be due to the inadequacy of the included 3NF. This is an old idea introduced 25 years ago \cite{Sla82} and a simple model with the Fujita-Miyazawa 3NF explained the difference between $a_{nn}$ obtained from two different configurations. Later study \cite{Tor93} reanalyzed data used in extracting $a_{nn}$ and found a much wider spread in $a_{nn}$ values and r3Bc showed that at specific production angle at each incident energy the 3NF effect vanish (for 13 MeV it vanishes around 45$^\circ$). The data \cite{Gon99} confirm small sensitivity to the 3NF. The difference in $a_{nn}$ cannot be due to MI as suggested \cite{Slo84}, since r3Bc showed that they are smaller than 0.2 fm \cite{Wit03}. Namely, MI influence both NN and 3N systems.

 9) Cross section and tensor analyzing powers Ayy at 19 MeV for the reaction H(d,pp)n in SCRE \cite{Ley06} - except for the coplanar star - do not agree with the predictions of r3Bc using TM, TM' and UIX or the coupled channel CD-Bonn + $\Delta$ and including Coulomb forces. Calculations were also done for N2LO including 3NF.

10) Radiative pd capture cross section, Axx and Ayy using polarized deuteron beams from RNCP at 140 and 200 MeV could not be explained \cite{Kud05} by r3Bc.

11) High precision NN and 3NF TM' and UIX could not explain 108 - 190 MeV $\sigma(\theta)$ data \cite{Erm03}.

12) Disagreement between $\sigma(\theta)$, Ay and polarization transfer coefficients Kyy' and Kxx' in the d-p elastic scattering at 250 MeV and all high precision NN + TM. The inclusion of TM 3NF deteriorates the fit to polarization transfer coefficients Kxz' and Kzx'. This is a first complete set of pd at intermediate energies \cite{Hat02}.

13) While most of the data from the reaction H(d,p)d at 135 MeV/nucleon \cite{Sek04} are explained well by including the 3NF, Axx, Ayy, Axz analyzing powers and Kyy' and Kxzy' could not be indicating inadequacy of the current 3NF.

14) Two-body $^3$He and $^4$He photodisintegration data measured at JLab at 0.35 - 1.5 GeV do not agree with the best available calculations, albeit these calculations are rather old-fashioned\cite{Ber06}.

15) Lithium problem - a statistically significant discrepancy between standard BBN predictions for $^7$Li and twice smaller observational value over a wide range of metallicities \cite{Pos06,Fie06}. Existence of the metastable (T $\geq$ 1000 s) negatively charged electroweak scale particle X$^-$ would alter the prediction of $^7$Li and other primordial abundances for A$\geq$5 via the formation of the bound state with nuclei during the BBN.
We might be tempted to repeat the famous statement that these discrepancies are 15 minor clouds on a bright sky, but we should know better.

\section{IS THERE AN END IN SIGHT FOR FEW BODY RESEARCH?}

	Successes described in II and III point toward a “Standard model” of FB theory. However, FB field is broader than few nucleon and there are many open issues: 

1) A hypernuclear event caused by cosmic rays was observed in 1953 \cite{Dan53}. In 70-ies kaon beams were developed. There are 35 known $ \Lambda$ hypernuclei from $^3_\Lambda$H to $^{209}_\Lambda$Bi mostly below A = 20. Only three double-lambda-hypernuclei $^6_{\Lambda  \Lambda}$He, $^{10}_{\Lambda  \Lambda}$Be and $^{13}_{\Lambda  \Lambda}$B are known \cite{Dan63}. Evidence for $^4_{\Lambda  \Lambda}$H was suggested \cite{Ahn01}. $ \Lambda  \Lambda$ hypernuclei provide information on $ \Lambda  \Lambda$ interaction, properties of multi-strange systems and neutron stars \cite{She06} and give insight into the onset $ \Lambda\Xi$ hypernuclei stability \cite{Fil02}. The $ \Lambda  \Lambda$ BE B$_{\Lambda\Lambda}$ was determined from $^6_{\Lambda  \Lambda}$He suggesting that effective $ \Lambda  \Lambda$ interaction is considerably weaker than thought earlier. Baryon-baryon interactions for N, $ \Lambda$, $\Sigma$ and $\Xi$ were studied by the Nijmegen group \cite{Nag79} and in chiral quark model \cite{Fer05}. Double $ \Lambda$ hypernuclei were studied in the relativistic mean-field theory and in the 4B model: x +$\alpha$ +$ \Lambda$ +$ \Lambda$, where x is n, p, d, t, $^3$He or $\alpha$.

             2) Renaissance of QCD spectroscopy. Proton energy distribution from the reaction 4He(K-stopped,p) shows a peak interpreted as a neutral 3 baryon state, M = 3117, $\Gamma<$ 21 MeV with isospin 1 and strangeness -1 and main decay mode NN$\Sigma$ \cite{Suz05}. The nature of $ \Lambda$ (1405) is not fully understood yet. It could be a 3q, a molecular-like meson-baryon bound state or exotic \cite{Zyc06}. The $\Sigma$(1480) is described as a bump by PDG \cite{Eid04}. The Crystal Ball study of the reaction $K^- p \rightarrow \pi^\circ\pi^\circ \Lambda$ do not show $\pi^\circ \Lambda$ resonance, but are dominated by $\Sigma^\circ$(1385) \cite{Pra04}. Recent study of the reaction $pp \rightarrow p K\pi X$, where X is the unidentified residue, is interpreted as evidence for a neutral hyperon resonance $Y^\circ$* decaying into $\pi^+X^-$ and $\pi^-X^+$ with M = 1480 $\pm$15 MeV, $\Gamma$ = 60$\pm$15 MeV and since it is neutral it can be either $\Sigma^\circ$ or $ \Lambda$. Light baryonic states were reviewed and search for $ \Lambda$*, $\Sigma$*, $\Xi$* and $\Omega$* states - expected but yet undiscovered - was stressed \cite{Nef02}. Relatively narrow deeply bound antikaon nuclear states in FB systems were suggested \cite{Aka02}and searched for \cite{Suz05}.
	The dominant problem is missing resonances - the link between genuine 3q resonances and scattering data \cite{Tho01}. Agreement on confidence level of important resonant states is still not reached. It is suspected that missing states couple weekly to the formation channels \cite{Cap00,Sva06}and it seems that it is related to the importance of inelastic channels as recently confirmed by the Zagreb group demonstrating that inelastic channels, particularly $\pi N\rightarrow\eta N$ are decisive in unambiguously confirming \cite{Cec06} one of controversial states: $N(1710)P_{11}$. Inelastic channel are very important and experimental efforts should be strengthened (proposals EPECUR and MIPP \cite{EPE05}). Mesons, baryons, resonances etc are topics of several conference series MENU, ETAMESON, NSTAR \cite{NST06} and of PSA workshops recently held in Abilene, Zagreb and Tuzla.

 	3) $\eta$ physics includes studies of $\eta$-mesic nuclei and of rare decays \cite{Nef05}. Charge conjugation invariance (CCI) forbids the decay of pseudoscalar mesons into an odd number of $\gamma$s and search for $\eta$ $\rightarrow$ 3$\gamma$ is a direct test of CCI for EM interaction of hadrons. The upper limit for the branching ratio (BR) for this process was established by two groups \cite{Nef05a}yielding BR $\leq$ 4$\times$10$^{-5}$ (Crystal Ball) and BR $\leq$ 1.6$\times$10$^{-5}$ (KLOE at DA$\Phi$NE). The process $\eta \rightarrow \pi^\circ\gamma\gamma$ is a test for $\chi$PT. Since there is no direct order $\gamma\pi^0$ coupling and the second order involves G-parity violating transition, the first contribution comes from the third order. In 1982 GAMS-2000 gave BR = (9.5$\pm$2.3) 10$^{-3}$, and if confirmed would be the first failure of $\chi$PT. Recent measurement gave BR = (3.2 $\pm$ 0.9) 10$^{-4}$, converted to $\Gamma$ = (0.45 $\pm$ 0.09$_{st}\pm$ 0.08$_{sy}$) eV \cite{Pra05}, vs. the $\chi$PT result of 0.42$\pm$ 0.20 eV. CBELSA TAPS collaboration is redoing this study with the aim of reaching a better accuracy \cite{Roy06}.

4) QCD allows systems with explicit gluonic degrees of freedom: hybrids (qq*g) and glueballs (gg). Hybrid configurations also appear as intermediate states \cite{Pop04}. Smoking gun for hybrids is to have JPC forbidden by qq* models, e.g. 0$^{--}$, 1$^{-+}$, 2$^{+-}$ \cite{Clo03}. Evidence for several hybrid candidates are found: i) 1$^{-+}$ in the reaction $p\pi^-\rightarrow \pi^-\pi^- \pi^+ p$ at 18 GeV/c in addition to expected states 1$^{++}$ a1(1260), 2$^{++}$ a2(1320) and 2$^{-+}$  $\pi_2$(1620) \cite{Ada98}, ii) states at M = 1709 and 2001 MeV and at 2096 MeV from the reaction $p\pi^-\rightarrow \pi^-\pi^-\pi^+\eta p$  \cite{Kuh04} and iii) exotic meson decays to $\omega$ $\pi^0$ $\pi^-$ \cite{Mlu04}. Evidence for glueballs are not found \cite{Kle04}. Strangeness in the proton and in N(1535) suggesting an important ss* component, relevant e.g. in the search for Cold Matter \cite{Eli01}, is discussed in \cite{Zou06,Tho06}.

5) There are speculations that the short range repulsion of the NN force is increased and the intermediate range attraction decreased in the medium \cite{Bey06,Yak05}.

6) Halo structure \cite{Jen04,Zhu93,Can06} (e.g. $^{14}$Be \cite{Lab01}), Borromean, tango and samba nuclei and quantum proximity resonances \cite{Hel96} are related to Thomas \cite{Tho35} and Efimov \cite{Efi70} effects. Borromean nucleus \cite{Yam05} is a 3B bound state where none of its 2B subsystems is bound, e.g. $^6$He. Excited state of $^{12}$C is a Borromean excited state of paramount importance in SINS. Tango nuclei have only one bound subsystem and in samba nuclei two of the 2B subsystems are bound. $^{10}$Be can be treated as a 3B ($^8$Be + n +n) or as a 4B problems, and there could be a molecular $\alpha$:2n:$\alpha$ band \cite{Fre06}. Using QFS and “inverse” kinematics to study the reaction H($^6$He, ab) and H($^8$He, ab), where a and b stand for pn, p$\alpha$ and p-$^6$He, momentum distributions inside $^6$He and $^8$He were measured \cite{Chu05}. 3B calculations are consistent with data on nn rms distances in $^6$He, $^{11}$Li and $^{14}$Be \cite{Tom05}. QF reactions (A(a+b) + B $\rightarrow$ C +D +b (initiated in 1967 by M. Furic \cite{Fur67}) and two spectators QFS: (A(a+b) + B(c+d) $\rightarrow$ C + D + bs + ds) (initiated by D. Miljanic in 1974), later studied at NRL \cite{Sla73,All78}, could be relevant for productions of neutron- and proton-rich nuclei. Studies of xH and xn  are initiated with $^{14}$Be and $^8$He beams. Resonant states in $^7$H are found by H(8He,pp)$^7$H, evidence that $^8$He contains $^6$He subsystem in an excited state 2$^+$ with a large fraction \cite{Mar02}and $^7$He levels were studied \cite{Boh01}. Similar studies are done for hypernuclei, e.g  $^7_{\Lambda\Lambda}$H \cite{Study}.

7) Session were devoted to atomic, molecular and condensed matter physics and chemistry at FB meetings starting from Nanning in 1985 \cite{Few85}. FB18 included talks on pairing of atomic fermions \cite{Hul06}, FB effects in cold atoms \cite{Ham06} and evidence for Efimov states in ultracold gas of Cs atoms \cite{Gri06}. Chemical potentials of the two spin states are equal in superconductors. Mismatched chemical potentials are important in magnetized superconductors and cold dense quark matter at the core of neutron stars. Pairing in a polarized gas of $^6$Li atoms was observed \cite{Par06}. Observation of Efimov states in nuclear systems is hampered by Coulomb forces and for A$\geq$4 there are no Efimov states \cite{Efi06,Ama73}. In molecular physics helium trimer is predicted to have an excited Efimov state but its existence is not confirmed \cite{Lim77,Bru05}. Studies of 3B recombination processes in ultracold gases with magnetically tunable 2B interaction of Cs atoms (thus varying 2B scattering length) established evidence for Efimov states \cite{Kra06}. The e$^-$-e$^-$ correlation in initial and final states of double ionization and (e,3e) processes is an important tool in atomic physics \cite{Gas06} and it persists even at high energy impacts as demonstrated by the asymptotic formula for the ratio of double to single cross section in Compton scattering\cite{Sur94}.
	
\begin{table}[ht]
\caption{$\chi^2$/datum for various NN potentials.}
\begin{tabular}{cccccccc}
\hline\hline
YEAR & N$^\mathrm{o}$ of data & Nijmegen PSA & CD-Bonn & AV18 & N3LO &N2LO& NLO\\
\hline
1992 & pp(1787) & 1.00 & 1.00 & 1.10 & & & \\
1999 & pp(2932) & 1.09 & 1.01 & 1.35 & & & \\
1999 & NN(5990) & 1.04 & 1.02 & 1.21 & & & \\
\hline
     & \multicolumn{2}{c}{0-290 MeV NN data}       &     &      & & & \\
\hline
1999 & np &      &     & 1.04 & 1.10 & 10.1 & 36.2 \\
1999 & nn &      &     & 1.38 & 1.50 & 35.4 & 80.1\\
\hline
     \multicolumn{2}{c}{N$^\mathrm{o}$ of parameters} & 35 & 38 & & 24 & 9 &  \\
\hline\hline
\end{tabular}
\end{table}

\begin{table}[ht]
\caption{Binding energies (MeV) of 3 $\leq$ A $\leq$ 16 calcualted with 3NF(37,37a)}
\begin{tabular}{cccccccccc}
\hline\hline
$^3$H & $^3$He & $^4$He & $^6$He & $^6$Li & $^7$Li & $^8$He & $^3$H & $^8$Be & $^{16}$O \\
\hline
Experiment & 8.488 & 7.72 & 28.296 & 29.27 & 31.995 & 39.2 & 31.598 & 56.5 & 127.619\\ \hline
AV18 & 7.628 & 6.917 & 24.07 & & 26.09 & & & &\\
F:CDB+TM & 8.478 & 7.735 & 29.25 & & & & & & \\
AV18+UIX & 8.478 & 7.733 & 28.34 & & & & & & \\
\hline
G:AV18+UIX & 8.487 & 7.739 & 28.33 & 28.1 & 31.1  & & 27.2 & 54.4 & \\
AV18+IL2   & 8.43  & 7.67  & 28.27 & 29.4 & 32.3  & & 31.3 & 56.6 & \\
\hline
N:AV8'+TM' &       &        &        & 28.189 &        & & & & \\
JISP16    & 8.496  & 7.797  & 28.374 & 28.32  & 31.00  & & & 53.3 & 133.8 \\
\hline
N:N2LO NN only &       &        &        & &        & 34.6 & & & \\
Full           &       &        &        & &        & 36.7-38 & & & \\
\hline\hline
\end{tabular}
\end{table}

\section{CHALLENGES}

      Surveys presented in V and VI show the need for extensive experimental studies - high accuracy and innovative. Physics depends on measurements and several proposals for new major facilities and upgrading of existing facilities were submitted. It is in decisions to build such facilities that planning of science comes as shown in a famous Bernal - Polanyi debate in the 40-ties. This is contained in a prescription for success: get the best people, give them best working conditions and do not disturb them.

Though most of discrepancies listed in V.3 were confirmed by more than one measurement, it is necessary to perform a systematic consistency check and pruning of FB data as was done for NN data paying attention to normalization errors. (There is currently a disagreement between KVI and RIKEN high energy Nd data \cite{Kal06,Sak06} and both data disagree with r3Bc.) Some data show clear evidence for 3NF and some agree better with no 3NF and with no Coulomb, though are affected by both. Likely, disagreements listed in V.3. do not imply any new physics, but demand careful inclusion of relativity, EM, 3NF, CD and CSB, possibly fine tuning of the NN force. Ay puzzle, n-$^3$H and ordering of levels of light nuclei suggest the need to fine tune P waves.
        High precision potentials from Table 1 are not phase equivalent, they predict different E$_t$ and have different P$_d$ : CDBonn predicts 8.013 MeV and 4.85\%, respectively, AV18 7.628 MeV and 5.76\%, Doleschall 8.48 MeV and 3.6\% and N3LO 7.855 MeV and 4.51\%. They have different short range interactions and differ conceptually. Moscow-Tübingen approach \cite{Fae05}introduces a dibaryon and claims that dibaryon degrees of freedom produce effects in FB systems and solve some puzzles of Nd systems \cite{Kuk06}. Diversities are rewarding, but it is imperative to understand the relationship among various high precision potentials, and multitude of NN potentials has to be only a temporary situation. Will EFT give an answer?

The first challenge is to obtain full nuclear interactions at N3LO (derivation of the 3NF in N3LO is in progress) and possibly even higher n. This implies understanding isospin violations and obtaining proper CD and CSB nuclear forces. Parameters in the EFT have to be uniquely determined and their physical meaning understood. 
       EFT was successfully used in several FB studies \cite{Kol99,Jen06}, some of relevance for BBN \cite{And05}. Attempts were made to use EFT to improve Shell model \cite{Hax01}. More studies within EFT $\chi$PT are required.
       It is necessary to perform rigorous 3B, 4B...  studies using NnLO and EM interactions. When we will have NnLO for n=4 and r3Bc, r4Bc etc done, then we will be able to conclude what are physical consequences when the calculations do not agree with the data. Therefore, a variety of several different approaches to calculate FB observables: momentum, configuration space, Faddeev and GFMC and NCSM should be encouraged. 
      	Relativistic effects were studied \cite{Wit05a} at En = 28, 65, 135 and 250 MeV including relativistic kinematics, boost effects and Wigner spin rotation and as an input a relativistic NN on shell equivalent to AV18. The subject was reviewed at this conference by F. Gross \cite{Gro06}. Relativity has to be fully and reliably taken into account at all levels: particle-particle and FB systems.
	The challenge is to obtain high precision baryon-baryon interaction. Progress is made by the J\"ulich \cite{Hai05} and Nijmegen \cite{Rij06} groups and discussed at this conference \cite{Hai06,Fer06}.
	FB studies are the ideal ground for searching for “new physics”. Examples are: i) studies of weak charge of the proton $QW(p) = 1 - \sin^2\theta_W$ at JLab planned to achieve 0.3\% accuracy of $\sin^2θ_W$, \cite{vOe06} and ii) test of chiral symmetry breaking in systems with strangeness. The DEAR (DAΦNE Exotic Atom Research) and SIDDHARTA (Silicon Drift Detector for Hadronic Atom Research by Timing Application) use precision X-ray spectroscopy of kaonic hydrogen and deuterium atoms to measure isospin dependent antikaon-N scattering lengths to test $\chi$SB \cite{Zme06}.
	FB research started with nuclear physics studies and - as a result of progress in theoretical studies and in technological development - a renaissance of nuclear physics is occurring. A special position of nuclear physics research at the intersection of particle, condensed matter and atomic physics and chemistry, biosciences and astrophysics makes it relevant now, when search for “new physics” is intensifying, when chemistry, atomic and nuclear physics employ ab initio procedures, when rare isotopes at the drip-line can be viewed as “femtoscience” and when astrophysics research critically depends on precise nuclear input. $^1$S$_0$ neutron superfluidity is relevant for phenomena occurring in the inner crust of the neutron stars. Precise $^3P_2-^3F_2$ pairing gap is important for cooling of neutron stars and cooling of young stars (T$\leq$10$^5$ y) is governed by neutrino emission (URCA process named after casino in Rio) \cite{Dea03,Jen06}. Studies of neutron and proton drip lines, of Efimov effects and halo nuclei are a high priority and RIB will find applications in life sciences.

Many topics, e.g. NN$\gamma$ and pentaquarks, and a wide spectrum of atomic and chemical studies are left out. I just gave a partial list of challenges facing our FB research community. All tasks and challenges facing FB research merge into a comprehensive task of building the knowledge-based society, where science and physics in particular play a crucial role. Physicists should be like an English admiral in the anecdote describing the visit of Peter the Great to England. Peter was so impressed by the English Navy that he told the English admiral ``If I were not the czar of all Russia, I would like to be an English admiral.'' to which the admiral replied ``Sire, if I were not an English admiral, my greatest wish would be to become an English admiral.''

\paragraph*{ACKNOWLEDGEMENT}
It would be impossible to write this paper without help and contributions of many of my colleagues and my special gratitude for their support and critical suggestions goes to my friends Calvin R. Howell, Ruprecht Machleidt, Kruno Pisk, Alfred \v Svarc, Lauro Tomio and Werner Tornow. I thank Mr. S. Ceci for his assistance. The support of the Ministry of Science of Croatia is acknowledged.



\begin{thebibliography}{aaaaaaaa}
\bibitem{Study} Study of 7$ \Lambda$H are performed at FINUDA.
\end{thebibliography}
\end{document}